\documentclass[5p,twocolumn,times,sort&compress]{elsarticle}

\usepackage{lineno,hyperref}
\usepackage{graphicx}
\usepackage{dcolumn}
\usepackage{bm}
\usepackage{comment}
\usepackage{rotating}
\usepackage{longtable}
\usepackage{float}
\usepackage{eucal}
\usepackage{csquotes}
\usepackage{xcolor}

\modulolinenumbers[5]

\begin{document}

\begin{frontmatter}







\title{\centering {Nuclear level density of $^{69}$Zn from gamma gated particle spectrum and its implication on $^{68}$Zn(n, $\gamma$)$^{69}$Zn capture cross section}}

\author[label1]{Rajkumar Santra}
\author[label2]{Balaram Dey}
\author[label3]{Subinit Roy\corref{cor1}}
\cortext[cor1]{Corresponding author at: Saha Institute of Nuclear Physics, Kolkata, India.}
\ead{subinit.roy@saha.ac.in}
\author[label4]{Md. S. R. Laskar}
\author[label4]{R. Palit}
\author[label3]{H. Pai}
\author[label5]{S. Rajbanshi}
\author[label1]{Sajad Ali}
\author[label1]{Saikat Bhattacharjee}
\author[label4]{F. S. Babra}
\author[label3]{Anjali Mukherjee}
\author[label4]{S. Jadhav}
\author[label4]{Balaji S Naidu}
\author[label4]{Abraham T. Vazhappilly}
\author[label4]{Sanjoy Pal}

\address[label1]{Nuclear Physics Division, Saha Institute of Nuclear Physics, 
and Homi Bhabha National Institute, Kolkata-700064, India.}
\address[label2]{Department of Physics, Bankura University, Bankura-722155, India.}
\address[label3]{Nuclear Physics Division, Saha Institute of Nuclear Physics, Kolkata-700064, India.}
\address[label4]{Department of Nuclear and Atomic Physics, Tata Institute of Fundamental Research, Mumbai-400005, India.}
\address[label5]{Deptartment of  Physics, Presidency University, Kolkata-700073, India.}

\begin{abstract}
Evaporated $\alpha$-spectra have been measured in coincidence with low energy discrete $\gamma$-rays from residual nucleus $^{68}$Zn populated in the reaction $^{64}$Ni($^9$Be,$\alpha$n)$^{68}$Zn at $E(^9$Be) = 30 MeV producing $^{73}$Ge compound nucleus. Low energy $\gamma$-gated $\alpha$-particle spectra, for the first time, have been used to extract the nuclear level density (NLD) for the intermediate $^{69}$Zn nucleus in the excitation energy range of E $\approx$ 5-20 MeV. The slope of NLD as a function of excitation energy for $^{69}$Zn matches nicely with the slope determined from RIPL estimates for NLD at low energies and the NLD from neutron resonance data. Extracted inverse NLD parameter (k = A/$\widetilde{a}$) has been used to determine the nuclear level density parameter value $a$ at neutron separation energy $S_n$ for $^{69}$Zn. Total cross section of $^{68}$Zn(n,$\gamma$) capture reaction as a function of neutron energy is then estimated employing the derived $a(S_n)$ in the reaction code TALYS. It is found that the estimated neutron capture cross section agrees well with the available experimental data without any normalization. The present result indicates that experimentally derived nuclear level density parameter can constrain the statistical model description of astrophysical capture cross section and optimize the uncertainties associated with astrophysical reaction rate.

\end{abstract}


\date{\today}
\end{frontmatter}

In the domain of statistical model description of nuclear reactions, level density as a function of excitation energy of the nucleus is one of the important quantities. A detailed knowledge of nuclear level density (NLD), especially in the region around the neutron separation of the nucleus, is crucial to understand various nuclear processes like nuclear fission \cite{capote}, multi-fragmentation \cite{bondorf}, spallation reaction \cite{davide} and capture reactions in nuclear astrophysics \cite{rauscher1, rauscher2}. 

In nuclear astrophysics, neutron capture reactions in $s$-and $r$-processes play a decisive role in the understanding of origin of elements heavier than iron. The description of neutron 
capture reactions relies on statistical model calculations to determine the astrophysical reaction rate. The Hauser-Feshbach model is used to calculate the reaction cross section. 
The model requires the optical model potentials (OMP) for particle transmission coefficients, nuclear level density (NLD) parameter and gamma ray strength function for extraction of $\gamma$-ray transmission coefficients \cite{Voinov}. The parameters of the model, however, are rather poorly constrained. One of the main sources of uncertainties in the predicted $n$-capture reaction comes from lack of experimentally determined NLD or due to lack of reliable description of NLD. An accurate description of NLD as a function of excitation energy and angular momentum is, therefore, essential in estimating the relevant reaction cross sections.   


A number of approaches have been proposed to understand NLD theoretically as well as experimentally. Theoretically, the NLD has been characterized by phenomenological analytical
expressions \cite{bethe,igna,gilbert} as well as calculations based on different microscopic approaches \cite{gori,grimes,ozen,hung1}. Experimentally, the NLD is estimated from
counting the levels at low excitation energy, from neutron resonance studies \cite{huiz}, by Oslo technique \cite{schiller}, using two-step cascade method \cite{bec}, by $\beta$-Oslo
method \cite{spy}, from $\gamma$-ray calorimetry \cite{ull}. These experimental techniques can only be used to extract the NLD up to the particle threshold energy and can be
extrapolated to higher energies using the functional form of the Fermi gas (FG) model \cite{bethe,igna,gilbert}. However, the functional form of NLD is not yet satisfactory
due to the lack of a trend in experimental data at high excitation energy and spin. Therefore, it is very important to acquire the experimental level density at low as well as high
excitation energy E$^*$ and spin J using different techniques. The particle evaporation technique is another approach to estimate the level density below as well as above the particle threshold energy \cite{voi1,bdey2,rami1,rami2,byun}. The validity of the particle evaporation technique has been checked in Ref. \cite{voi1}, where it has been shown that particle spectra are most suitable for NLD studies. It should be mentioned that although the particle evaporation technique is model dependent, it can be used to know the exact trend of experimental NLD as a function of E$^*$ with a proper normalization based on the density of known discrete levels and the average neutron resonance spacing which are generally well documented in literature \cite{capote,mugh}. In addition, the inverse level density parameter ($k$ = A/$\widetilde{a}$), which is an important ingredient in the functional form of
FG model \cite{bethe}, should be measured experimentally as it varies strongly with temperature (T) and spin (J) \cite{shlomo,bdey3,kban}. The potential problem with the particle evaporation technique is that multistep and direct reactions may contribute. Recently, low-energy light-ion beams (d, $\alpha$) have been used to extract the NLD from a particular channel and the contribution from direct reaction has been ruled out by measuring the particle spectra at backward angles and the angular distributions of the particles \cite{bdey1,voi1,bdey2,rami1,rami2,byun}.

A new approach has been adopted for the first time in the present study. We have measured the particle evaporation spectra gated by low energy $\gamma$-rays decaying from states 
that can be populated predominantly through compound nuclear (CN) reaction process eliminating the possible contributions from multistep and direct reactions in the outgoing particle spectra. The selection of these particular de-excitation $\gamma$-rays of residual nucleus provides us the scope for extraction of the required NLD for a particular nucleus after particle evaporation.

In this work, $\gamma$-gated $\alpha$-particle spectrum has been used to extract the inverse level density parameter $k$ of FG model and the NLD in the energy range of E $\approx$ 4-20 MeV for the nucleus $^{69}$Zn produced through {\it $\alpha$n} decay channel in the reaction $^{9}$Be + $^{64}$Ni populating the compound nucleus $^{73}$Ge. The obtained level density are normalized to the density of known levels in the discrete energy region \cite{RIPL3}as well as to the level densities at the neutron separation energy \cite{NLD_Bn} in order to understand the exact trend of experimental Nods as a function of excitation energy. Subsequently, the extracted $k$ value has been used as an input parameter in FF model prescription to calculate the cross sections of 
$^{68}$Zn(n, $\gamma$)$^{69}$Zn capture reaction. 
\begin{figure}
\begin{center}
\includegraphics[height=7.0cm, width=9.4cm]{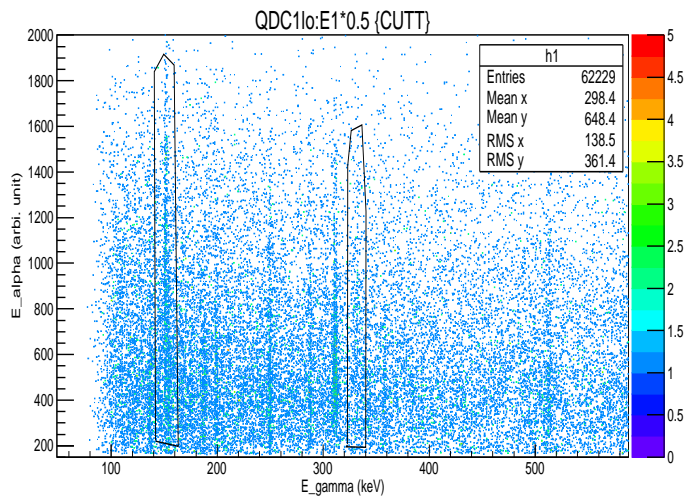}
\caption{\label{fig1} (Color online) $\alpha$-$\gamma$ coincidence matrix extracted from raw particle-$\gamma$ matrix. The events of interest are bounded by solid red boxes.}
\end{center}
\end{figure}

\begin{figure}
\begin{center}
\includegraphics[height=6.5cm, width=9.0cm]{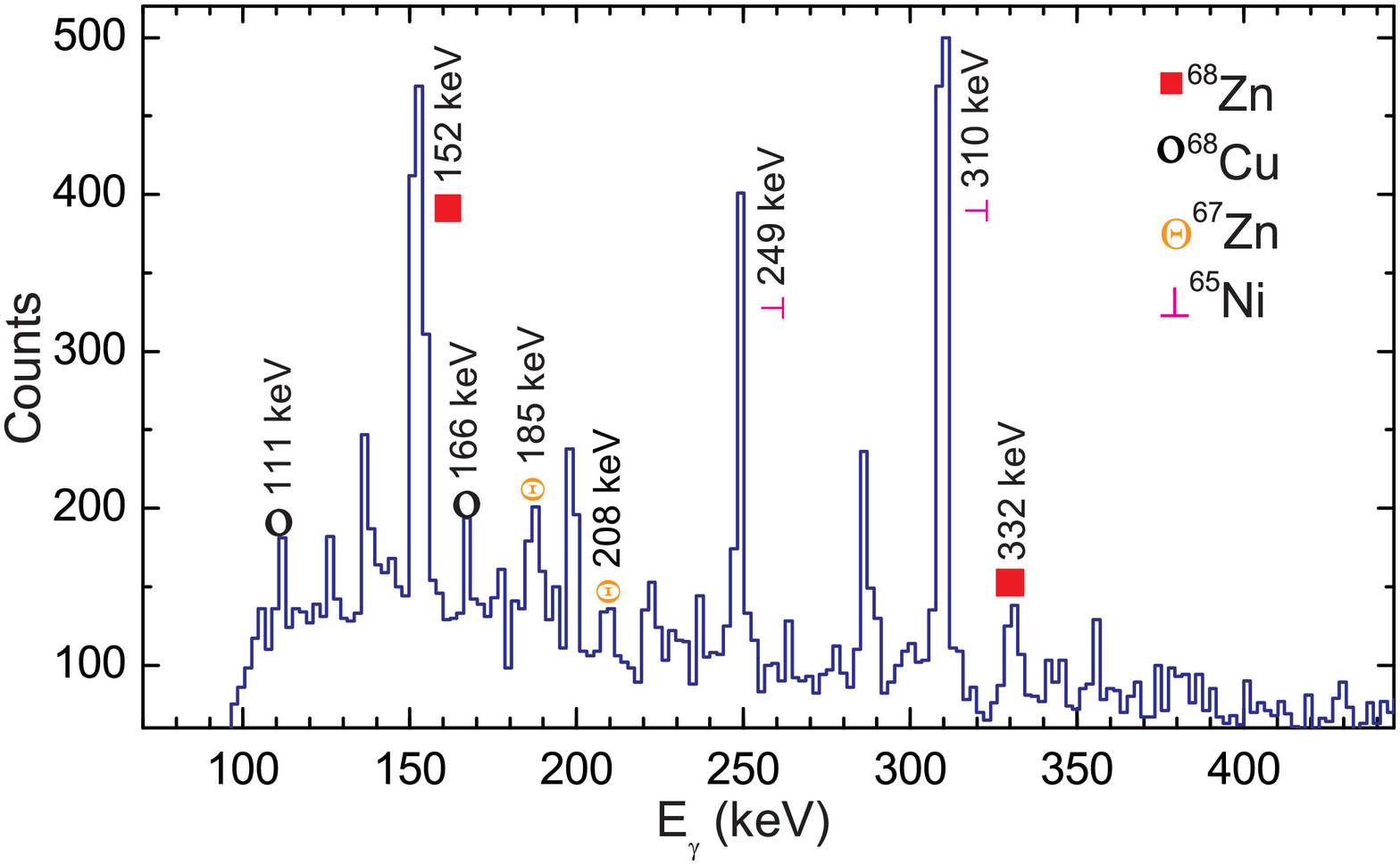}
\caption{\label{fig2} (Color Online) Projected gamma energy spectrum from Fig. \ref{fig1}. Symbol with $\gamma$-energy indicate $\gamma$-lines of different residual nuclei associated 
with $\alpha$-emitting channels. Gamma energies of $^{68}$Cu \cite{Tikku}, $^{67}$Zn \cite{Wender} and $^{65}$Ni \cite{Cochavi} are confirmed from $\gamma - \gamma$ coincidence matrix. }
\end{center}
\end{figure}

The experiment was performed at NARC-RIFT Pellet Lina Facility in Mumbai, India. A self-supporting $\approx$99 $\%$ enriched metallic $^{64}$Ni target of thickness about 500 $\mu$g/cm$^2$ thickness from Oak Ridge National Laboratory, USA, was used for the present experiment. The target was bombarded by $^9$Be beam at 30 MeV populating $^{73}$Ge at 41.8 MeV excitation energy. Eight CsI(Tl) detectors, each of thickness 3 mm (size 15x15 mm$^2$), were used to detect the outgoing charged particles. Two sets of four detectors each were placed symmetrically about the beam axis at 5 cm from the target center. The detectors were put on  both  sides  of  the  beam  line  covering  an angular  region  from  22$^\circ$ to 67$^\circ$ in  the  reaction plane. Tantalum absorbers of thickness 30 mg/cm$^2$ were used before the CsI(Tl) detectors to stop the elastically scattered particles from entering the detectors. CsI(Tl) detectors were calibrated using $^{229}$Th source. De-exciting $\gamma$-rays  of  residual  nuclei were detected  using the $\gamma$-detector setup consisting of 14 Compton-suppressed Clover detectors placed at 40$^\circ$, 90$^\circ$, 140$^\circ$, 115$^\circ$and 157$^\circ$ with  respect  to  the  beam  direction.  Data  were recorded in list mode in a digital data acquisition system  (DDAQ)  based  on  Pixie-16  modules  of XIA-LLC, which provided both energy and timing information. Data  were  sorted using the  Multiparameter time stamped based Coincidence  Search program MARCOS \cite{palit} to generate the particle-gamma matrix file. 
Coincidence $\alpha$-$\gamma$ events are sorted into a matrix with the $\alpha$-energy E$_\alpha$ versus the $\gamma$ energy E$_\gamma$ as shown in Fig. \ref{fig1}. The projected $\gamma$-spectra of E$_\alpha$ versus E$_\gamma$ matrix is shown in Fig. \ref{fig2} and the various transitions of different $\alpha$ channels have been identified. 

\begin{figure}
\begin{center}
\includegraphics[height=9.5cm, width=9.5cm]{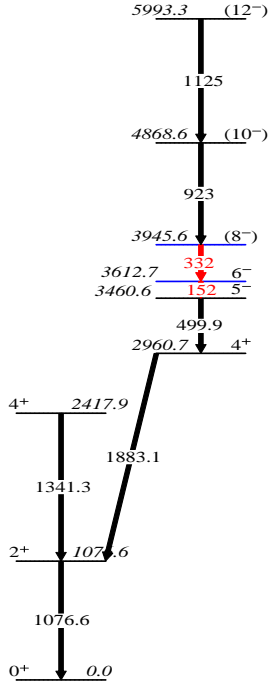}
\caption{\label{fig3} (Color Online) Partial level scheme of $^{68}$Zn based on its population in the present experiment. Transitions of interest and states shown in red and blue respectively. }
\end{center}
\end{figure}
\begin{figure}
\begin{center}
\includegraphics[height=9.0 cm, width=8.0 cm]{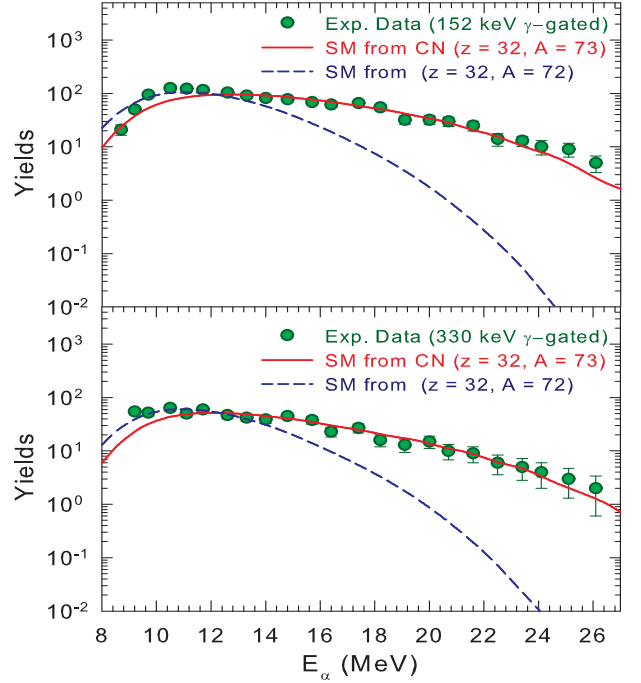}
\caption{\label{fig4} (Color online) Filled symbols represent the experimental $\gamma$-gated alpha energy spectra. Lines represent the statistical model calculations with CASCADE code. Red continuous line represents spectrum of first chance $\alpha$ decay from compound nucleus $^{73}$Ge. Black dashed line represents the contribution of $\alpha$ emission 
following first step one neutron decay of $^{73}$Ge. }
\end{center}
\end{figure}

While extracting the NLDs from particle evaporation spectrum, it is to be ensured that the contributions from non-compound processes are negligibly small. Another necessary condition for extracting the NLDs in this manner is that the particle spectrum be from first-chance emission. These are the pre-requisites for the extraction of NLD parameter by particle evaporation technique. To select out purely compound events, the $\gamma$-decay (E$_{\gamma}$ = 332 and 152 keV of $^{68}$Zn) of lowest lying negative parity states 6$^-$ and 8$^-$ have been chosen to gate the $\alpha$-spectrum. It should be mentioned that the level scheme of $^{68}$Zn (as shown in Fig. \ref{fig3}) from the $\gamma$-$\gamma$ coincidence matrix of present experiment also conforms the level scheme reported earlier in Ref. \cite{Neal}.
The nucleus $^{68}$Zn can be produced directly by $\alpha n$ decay of compound nucleus $^{73}$Ge or by $n$ emission after incomplete fusion/transfer of $^5$He fragment from $^9$Be to $^{64}$Ni. Again, $^{68}$Zn can also be produced by direct transfer of $\alpha$-fragment of $^9$Be to the excited states of $^{68}$Zn. The CN decay can populate the 6$^-$ and 8$^-$ states in $^{68}$Zn residue. Direct $\alpha$(0$^+$) transfer to $^{64}$Ni(0$^+$) can not populate these even spin, odd parity states. In case of transfer of heavier $^5$He fragment having Q-value of +10.24 MeV, the outgoing $\alpha$ from $^9$Be will have kinetic energy in the range of 36 to 39.44 MeV within the measured angular domain of 22$^\circ$ to 67$^\circ$. On the other hand, kinematically the energy of break up $\alpha$ corresponding to incomplete fusion of $^5$He with $^{64}$Ni target will lie within 8 to 12 MeV. Thus, in the measured $\gamma$-gated $\alpha$-particle energy spectrum, the contributions of different reaction channels, other than the CN process, beyond 12 MeV kinetic energy will be negligibly small. The $\gamma$-gated $\alpha$-energy spectra are shown in the upper and lower panels of Fig.\ref{fig4}. The $\gamma$-gated alpha energy spectra are converted into the CM frame in order to compare them with the statistical model calculations and carried out to investigate the NLDs as a function of excitation energy.          
The statistical model calculations (CASCADE) \cite{cas} have been carried out to fit the 152 and 332 keV $\gamma$-gated $\alpha$ spectra. The FG model \cite{bethe} of nuclear level density has been used in CASCADE code and is given by
\begin{equation}
\rho(E^*,J) = \frac{2J + 1}{12\theta^{3/2}} \sqrt{a} \frac{exp(2\sqrt{a U})}{U^{2}}
\end{equation}
where, U$=$E$^*$$-$$\frac{J(J+1)}{\theta}$$-$S$_\alpha$$-$$\Delta$P, E$^*$ being the excitation energy of the compound nucleus, $\theta$$=$$\frac{2I_{eff}}{h^2}$, with I$_{eff}$, S$_\alpha$ and $\Delta$P being the effective rigid-body moment of inertia, $\alpha$ separation energy and pairing energy, respectively. Ignatyuk prescription \cite{igna2} of level density parameter $a$, which takes into account the shell effects as a function of excitation energy is adopted and it is expressed as 
\begin{equation}
a = \tilde{a}[1 + \frac{\delta S}{U} [1-\exp(-\gamma U)]] 
\end{equation}
where, $\tilde{a}$ = A/$k$ and $k$ is inverse level density parameter. $\delta S$ is ground-state shell correction defined as the difference of the experimental and theoretical (liquid drop) masses. $\gamma^{-1}$ = $\frac{0.4A^{4/3}}{\tilde{a}}$ is the rate at which the shell effect is damped with the increase in excitation energy. The optical model potential parameters for alpha transmission coefficient are taken from Refs.\cite{OPM}. The moment of inertia of the CN is taken as $I_{eff} = I_0(1 + \delta_1 J^2 + \delta_2 J^4$), where $I_0$(= $\frac{2}{5}$$MA^{5/3}r_0^2$) is the moment of inertia of a spherical nucleus, $\delta_1$ (= 2 $\times$ 10$^{-6}$) and $\delta_2$ (= 2 $\times$ 10$^{-8}$) are the deformability parameters, r$_0$ is the radius parameter and $J$ is the total spin of the nucleus. The effect of the deformability parameters $\delta_1$ and $\delta_2$ has been checked and found to be insignificant. The shape of $\alpha$ energy spectra depends mostly on the level density parameter and partly on the potential parameters. The normalized fits from statistical model calculation are shown in Fig.\ref{fig4} in comparison with the data. The solid curve is the prediction for first chance $\alpha$ evaporation from the compound nucleus with a subsequent $n$ emission while the dashed line represents
the prediction for first chance $n$ evaporation followed by $\alpha$ emission. Same normalization has been used for both the curves. It is clear from the Fig.\ref{fig4} that ($n\alpha$)
decay has insignificant contribution compared to ($\alpha n$) decay channel above 12 MeV $\alpha$ energy. Thus the value of $k$ has been extracted from the best-fit statistical model calculations using a $\chi^2$-minimization in the energy range of E$_{\alpha}$ $\approx$ 12-26 MeV. The quality of over all statistical model fit to the $\alpha$ evaporation spectra also establishes the fact that the contribution of pre-equilibrium $\alpha$ emission, if any, in the higher energy region is insignificant and does not affect the extraction of
inverse density parameter. The extracted values of inverse level density parameter ($k$ = $A/\tilde{a}$) are 9.5 $\pm$ 0.6 MeV and 9.7 $\pm$ 0.6 MeV from 152 keV and 
332 keV $\gamma$-gated $\alpha$ spectra, respectively. 
\begin{figure}
\begin{center}
\includegraphics[height=5.5 cm, width=8.0 cm]{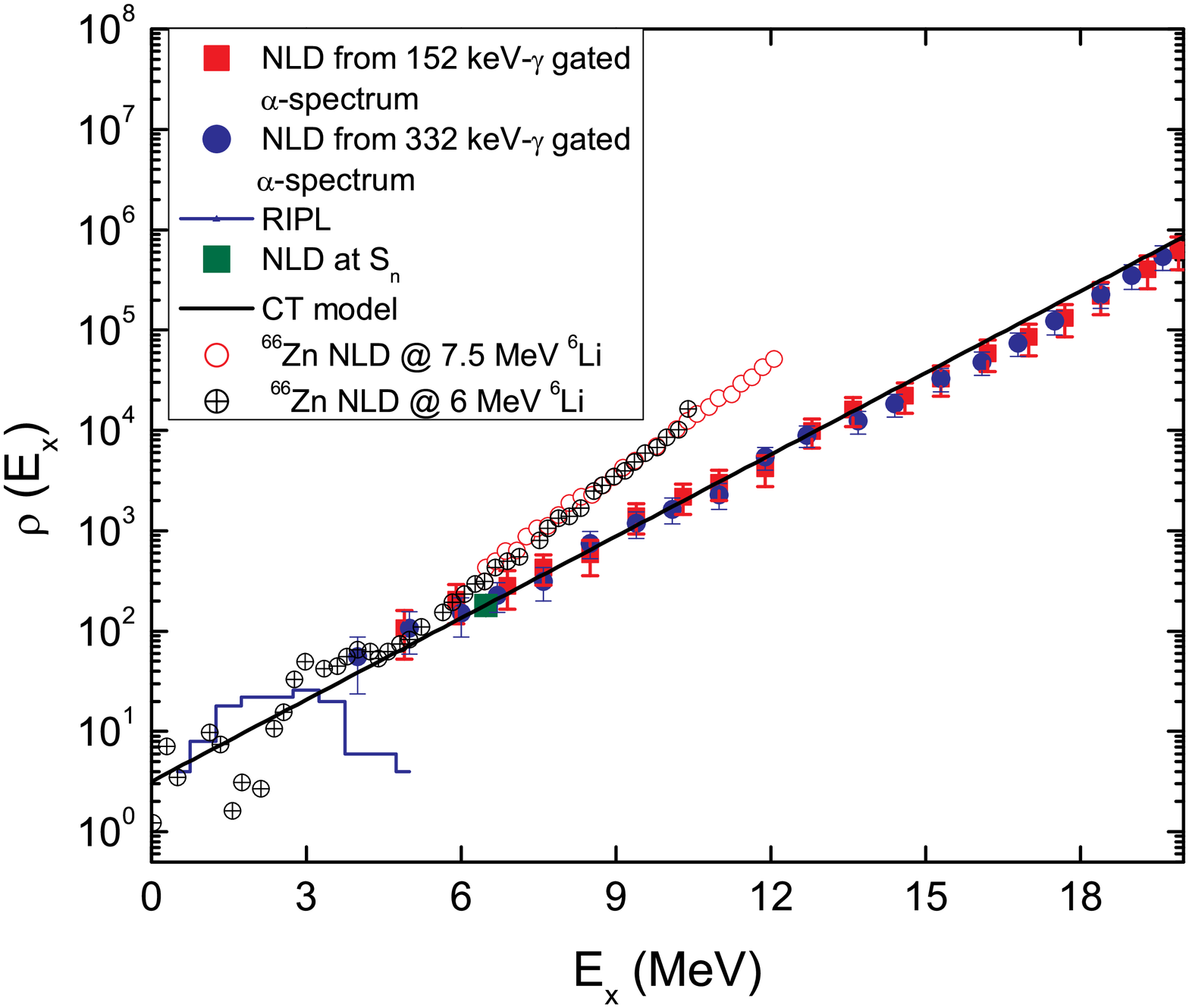}
\caption{\label{fig5} (Color online) Nuclear level densities as a function of excitation energy. Histogram represents the NLD taken from RIPL3, filled symbols represent the present extracted NLD from $\gamma$-gated alpha spectra, filled square (green) denotes the NLD from neutron resonance data determined at neutron binding energy \cite{NLD_Bn}. Data for $^{69}$Zn 
from present experiment is compared with the NLD of $^{66}$Zn nucleus \cite{rami1}}
\end{center}
\end{figure}

Finally, the experimental level density of residual nucleus has been determined in terms of the measured and calculated yields of $\alpha$ emission following the 
prescription of Refs.\cite{voi1,rami1, rami2, drc1995}
\begin{equation}
\rho_{\rm exp}(E_{X}) = \rho_{\rm fit}(E_{X}) \frac{(d\sigma/dE)_{\rm exp}} {(d\sigma/dE)_{\rm fit}} ~. 
\end{equation}
here, $(d\sigma/dE)_{\rm exp}$ and $(d\sigma/dE)_{\rm fit}$ are proportional to the experimental and the best-fit theoretical $\alpha$ energy spectra, respectively. $E_{X}$ = $U - E^{\rm CM}_{\rm \alpha}$ is the effective excitation energy, where $E^{\rm CM}_{\rm \alpha}$ is the alpha energy in the center-of-mass frame.  It should be pointed out that the state with maximum angular momentum in the level scheme of $^{68}$Zn is found to be 12 $\hbar$ in the present reaction. Therefore, the angular momentum range of 152 keV and 332 keV $\gamma$-gated particle spectra are typically $\approx$ 6-12 $\hbar$ and $\approx$ 8-12 $\hbar$, respectively. Here, the mean angular momenta (J = 9$\pm$3 and 10$\pm$2 $\hbar$ 
for 152 keV and 332 keV $\gamma$-gated alpha spectra, respectively) have been considered in the calculation of level density.
\begin{figure}
\begin{center}
\includegraphics[height=5.5 cm, width=8.0 cm]{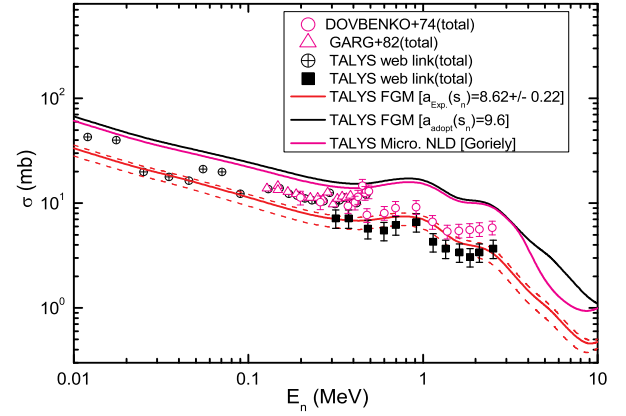}
\caption{\label{fig6} (Color online) $^{68}$Zn(n, $\gamma$)$^{69}$Zn capture cross-section as function of neutron energy. Reported experimental data are compared with the results obtained from TALYS calculation using the present experimentally measured level density parameter.}
\end{center}
\end{figure}
The level density of $^{69}$Zn residual nucleus as a function of excitation energy is shown in Fig.\ref{fig5}. The uncertainty in the level density due to statistical model parameters has been checked and found to be $\approx$ 10\%. The extracted level density has been normalized with respect to the level density at the neutron separation energy from neutron resonance data \cite{NLD_Bn} and it is seen that the slope of the normalized NLD nicley mathces with the level density of known levels in the discrete energy region taken from RIPL-3 \cite{RIPL3}. Egidy {\it et al.} \cite{NLD_Bn} studied the systematic behaviour of nuclear level density parameters of 310 nuclei. In their systematic study, they reported that the NLDs for even-mass nuclei at neutron separation energy $S_n$ are much higher (few orders of magnitude) than their odd-mass isotopes. To check this systematic behavior, we have shown in Fig. \ref{fig5} the available NLD data of $^{66}$Zn from Ref. \cite{rami1} measured using neutron evaporation spectrum. The systematic behaviour is clearly seen for Zn isotopes.

The importance of this extracted NLD lies in the measurement of inverse level density parameter ($k = \tilde{a}/A$) from
evaporated alpha energy spectra gated by the chosen low energy discrete $\gamma$-rays. In addition, the extracted NLD have been compared with constant temperature (CT) formula $\rho_{\rm CT}$ = $\frac{1}{T}$e$^{(E-E_0)/T}$. The value of E$_0$ and T are found to be -4.15 MeV and 1.77 MeV, respectively, which nicely explain the present extracted NLD 
as shown in Fig. \ref{fig5}. It should be pointed out here that the obtained E$_0$ and T do not corroborate with the systematic values reported in Ref.\cite{iwamoto}. In extracting these parameters the authors probably considered only the RIPL data. The present E$_0$ and T values are obtained by taking into consideration the NLD at neutron separation energy from neutron resonance data \cite{NLD_Bn}, the low energy (up to 7 MeV excitation energy) data of present measurement and the RIPL data \cite{RIPL3}. The value of T resulted 
from CT model fit also corroborates with the T = 1.91 determined from the relation T$=\sqrt{\frac{kU}{A}}$ \cite{voinov06} where U=E$^*$-E$_{rot}$-$\delta_P$-S$_{\alpha}$. The CT model fit describes the higher energy data upto 20 MeV within the error bar.

Furthermore, the extracted inverse level density parameter $k$ is used to calculate the nuclear level density parameter $a$ at neutron separation energy $S_n$ from Eq.2 
for $^{69}$Zn. Taking $U(S_n) = S_n - \Delta =$ 5.114 MeV determined from $S_n$ = 6.48 MeV and $\Delta$ = 1.368 MeV \cite{nndc} and $\delta S$ = 3.37 MeV from Ref.\cite{moeller}, we obtained $a(S_n)$ = 8.625 $\pm$ 0.225. The value has been utilized subsequently in TALYS1.9 nuclear reaction code \cite{talys} to calculate $^{68}$Zn(n, $\gamma$)$^{69}$Zn capture cross-section.The Fermi gas model of NLD \cite{bethe} has been used in the TALYS code. The neutron capture process is predominantly of $E1$ type, only the $E1$ strength function has been considered taking a generalized Lorentzian form \cite{kopecky}. This generalized Lorentzian form is also constrained by using the present level density parameter $a(S_n)$. So in the TALYS calculation we could constrain both NLD and $\gamma$sF with the use of $a(S_n)$ from the present maesurement. Global neutron optical model potential valid over the energy region of 0.001 MeV to 200 MeV for the mass range of 24$\le$A$\le$209 as the neutron potential in the statistical model calculation \cite{koning}. In the code, all other parameters are kept fixed except the nuclear level density 
parameter, wwe have measured the particle evaporation spectra gated by low energy $\gamma$-rays decaying from states that can only be populated through compound nuclear (CN) reaction.hich is taken from the present measurement. Interestingly, it is observed that the reaction cross-sections obtained from TALYS calculation using the measured NLD 
parameter explain the available data quite nicely without any further normalization, as shown in Fig. \ref{fig6}. The estimation with the systematic value of NLD parameter 
(a($S_n$)=9.614) clearly over predicts the data in the energy region of measurement.

In summary, the low energy (E$_{\gamma}$ = 152 and 332 keV) $\gamma$-gated alpha emission spectra from the reaction $^{9}$Be + $^{64}$Ni have been measured. The $\gamma$-gated alpha energy spectra is predominantly from the compound nuclear events ensured by the even spin, odd parity (6$^-$ or 8$^-$) of the decaying states in $^{68}$Zn, the residual nucleus from
$\alpha n$ decay of compound nucleus $^{73}$Ge. The measured alpha energy spectra have been compared with the statistical model calculations to extract the NLD parameter and utilized to extract the level density of $^{69}$Zn nucleus as a function of excitation energy. The obtained NLD parameter evaluated at neutron separation has been used in TALYS code to calculate the $^{68}$Zn(n, $\gamma$)$^{69}$Zn capture cross-sections. Excellent agreement with measured (n, $\gamma$) cross section, does highlight the objective of experimentally constraining 
the parameters of statistical model for more accurate description of astrophysical reactions. 
 
\begin{center}
$\textbf{Acknowledgements}$
\end{center}
Authors thank the BARC-TIFR PLF staff for uninterrupted, steady  beam during the experiment.  We would also like to thank Prof. G. Gangopadhyay, Calcutta University and Dr. D. Pandit, 
VECC for their help and advice in this project. Author H. Pai is grateful for the supprot of the Ramanujan Fellowship Research Grant under SERB-DST (SB/S2/RJN-031/2016), Govt. of India. This work is supported by Department of Atomic Energy, Government of India (Project Identication Code: 12-R$\&$D-TFR-5.02-0200).

\end{document}